Deriving the factor endowment–commodity output relationship for Thailand (1920-1927) using a three-factor two-good general equilibrium trade model

Yoshiaki Nakada, Faculty of Agriculture, Kyoto University, Kitashirakawa-oiwake-cho, Sakyo, Kyoto, Japan 606-8152

E-mail: nakada@kais.kyoto-u.ac.jp

Abstract

Feeny (1982, pp. 26-28) referred to a three-factor two-good general equilibrium trade model, when he explained the relative importance of trade and factor endowments in Thailand 1880-1940. For example, Feeny (1982) stated that the growth in labor stock would be responsible for a substantial increase in rice output relative to textile output. Is Feeny's statement plausible? The purpose of this paper is to derive the Rybczynski sign patterns, which express the factor endowment–commodity output relationship, for Thailand during the period 1920 to 1927 using the EWS (economy-wide substitution)-ratio vector. A "strong Rybczynski result" necessarily holds. I derived three Rybczynski sign patterns. However, a more detailed estimate allowed a reduction from three candidates to two. I restrict the analysis to the period 1920-1927 because of data availability. The results imply that Feeny's statement might not necessarily hold. Hence, labor stock might not affect the share of exportable sector in national income positively. Moreover, the percentage of Chinese immigration in the total population growth was not as large as expected. This study will be useful when simulating real wage in Thailand.

Keywords: three-factor two-good model, general equilibrium, Rybczynski result, factor endowment–commodity output relationship, EWS (economy-wide substitution)-ratio vector, Chinese immigration, Thailand

1. Introduction

Feeny (1982, pp. 26-28; Appendix 2, pp.162-170) referred to a three-factor two-good neoclassical model (hereinafter, $3 \times 2$ model), when he explained the relative importance of trade and factor endowments in Thailand 1880-1940 (see also Heady (1983, p. 195), who appreciated Feeny with regard to this reference). Feeny (1982, p. 26), stated, "General equilibrium trade models are an efficient analytical tool with which to examine the relationships between the trends in output prices, input stocks, and relative factor prices." His model included two goods (rice and textiles) and three factors (land, labor, and capital), where land is specific to agriculture. We can consider this model as a limiting case of factor intensity (see Batra and Casas (1976, pp. 26-27)) (hereinafter, BC) and Bliss (2003, p. 268)). We call this type of model an asymmetrical $3 \times 2$ model.

Feeny (1982, p. 26) stated, "This model is a reasonable description of the Thai economy in the period under review." In Appendix 2, Feeny (1982, pp. 169-170) referred to the equations derived by Hueckel (1972, Chapter 2) in his thesis, which was later published as a book (see Hueckel (1985)). The equations that Feeny referred to appear in Hueckel (1985, Chapter 2, pp. 66-79). Feeny (1982, p. 27) also referred to Hueckel's results in Table 3-16. Feeny (1982, p. 169)) stated, "Table 3-16 presented part of Hueckel's Table 1 [see Hueckel (1985, p. 72, Table 1)]. The full table is presented below as Table A2-3." Apparently, Hueckel derived his results in these tables from his equations.

However, upon reviewing Feeny's work, I detected an error in his statement. For example, the equations derived by Hueckel (1985, p. 68-69, note 18) include a serious mistake. Hueckel used elasticity of substitution defined for two factors in both sectors, whereas he should have employed Allen's partial elasticity of substitution in sector 1 because sector 1 employed three factors.

Hueckel published a part of his work as a journal article (see Hueckel (1973)), which Feeny (1982) never referred to. In his explanation about the concept of "elasticity of substitution," Hueckel (1973) explicitly refers to this error.[1]

Therefore, the question arises: Is Feeny's statement plausible? Notably, Feeny (1982, p. 28) stated, "Equation 12 in Appendix 2 [originally derived by Hueckel (1972)] indicates that the terms of trade that favored agriculture [in Thailand] would be crucial in explaining the gains made by rents

---

[1] According to Hueckel (1973, p. 377), $\sigma_{N,L}$ and $\sigma_{K,L}$ are the elasticities of substitution between land and labor and between land and capital in agriculture, respectively, and $\sigma_M$ is the elasticity of substitution between labor and capital in manufacturing. Specifically, Hueckel (1973, p. 377, note 26, eq. (d)) presented the equations shown below for his "partial" elasticity of substitution.

$$\sigma_{N,L} = \frac{a_{NA}{}^* - a_{LA}{}^*}{R_L{}^* - R_N{}^*} \quad \sigma_{K,L} = \frac{a_{KA}{}^* - a_{LA}{}^*}{R_L{}^* - R_K{}^*} ; \sigma_M = \frac{a_{KM}{}^* - a_{NM}{}^*}{R_N{}^* - R_K{}^*}, \qquad (a)$$

where $a_{ij}$ denotes "input coefficients..., which represent the amount of the $i$th factor necessary to produce one unit of the $j$th output." $R_i$ "represents the value of marginal product of the $i$th input." $i = L, K, N, j = A, M. L, K,$ and $N$ denote land, capital, and labor, respectively. $A$ and $M$ denote agriculture and manufacture, respectively (Hueckel (1973, p. 375). Apparently, this equation is similar to those presented by Jones (1965, p. 560, eqs. (8)-(9)). Hueckel (1973, p. 377 note 26) continued, "The question of the formulation of elasticities of substitution, when the production function contains three inputs, is a difficult one. The formulations employed here for the agricultural sector simplify the algebra greatly and should be interpreted as 'partial' elasticities of substitution, which describe the change in the marginal products of the two designated factors resulting from an alteration in those two factors in such a way as to hold the third factor and output constant. It can be shown that in the special case of the Cobb–Douglas production function with constant returns to scale these substitution elasticities are all unity, a fact which will be useful below." After all, Hueckel (1973, p. 386) stated that he employed that assumption.

over wages." He continued, based on Table 3-16, "The growth in the terms of trade and the growth in the labor and land stocks would be responsible for the large growth in rice output relative to textile output which occurred."

In other words, Feeny thought he had succeeded in explaining:

(i)   How terms of trade affect rents over wages,
(ii)  How terms of trade affect rice output relative to textile output, and
(iii) How factor endowment affects rice output relative to textile output.

Feeny did not analyze the data on textile output at all, and his explanations are not self-evident. In mathematical terms, for example, Feeny's statement that the growth in labor stock would be responsible for relative output growth implies that

$$(X_1/X_2)^*/V_L^* = X_1^*/V_L^* - X_2^*/V_L^* = (+) . \qquad (1)$$

where $X_j$ denotes the amount of good $j$ produces $(j=1,2)$, $V_i$ is the supply of factor $i$ $(i=T,K,L)$ $T$, $K$, and $L$ refer to land, capital, and labor, respectively. The asterisk denotes the rate of change (e.g., $X_j^* = dX_j/X_j$). However, to my knowledge, other than Nakada (2016, Appendix B), no one has analyzed a sufficient condition for the left-hand side of (1) to be positive. In order to prove that this condition holds, some assumptions on parameters need to be made.

Feeny did not state the reason why he explained the relative output growth. Equation (A.6) implies that if the labor stock affects the growth of rice output relative to textile output positively, it simultaneously affects the share of exportable sector in national income positively.[2]

To the best of my knowledge, other than Feeny (1982) and Hueckel (1973, 1985), Yohe (1979), Daniels *et al.* (1991) and Bliss (2003) alone referred to the asymmetrical 3 × 2 model (see Yohe (1979, p. 188), Daniels *et al.* (1991, p. 249), and Bliss (2003, p. 268, p. 274)).[3] However, it is hard for us to understand these studies. Bliss analyzed some basic relationships in the model and tried to apply them to British economic history. However, he did not present the process of computation and solutions. On the other hand, Daniels *et al.* (1991, pp. 252-253, Table 2) analyzed the response of the local economy to an 18% decrease in the price of wood products assuming full employment. However, Daniels *et al.* (1991, pp. 248-250) did not present the entire computation. Because Yohe's model referred to the basic equations in BC, we can consider it as the modified version of BC's 3 x 2 model. His equations are complicated. Thus, I am uncertain as to whether all their conclusions are plausible (however, these points are not discussed in this paper).

---

[2] With regard to agriculture's share, Hueckel (1973, p. 394) only stated, "First, the pressures created by the war on output prices and factor supplies contributed to the increase in agriculture's share of total output which can be seen in the data from this period." However, he did not show the change in agriculture's share using mathematical expression. Agriculture was an importable sector in Britain during 1793-1815, as Hueckel (1973, p. 368) stated.

[3] The model in Yohe (1979, p. 188) included three factors (capital, labor, and pollution) and two sectors (a polluting sector and a nonpolluting sector) where pollution is specific to the polluting sector. The model in Daniels *et al.* (1991) included three factors (capital, labor, and stumpage) and two goods (wood products and generalized all-other-goods). The model in Bliss (2003) included three factors (land, capital, and labor) and two goods (agriculture and manufacturing).

Here, the following question arises. Why do we need to analyze the asymmetrical $3 \times 2$ model? It seems very complicated. Only a few researchers have analyzed it. For example, if we assume the value of $\theta_{T2}$ is small enough, say,

$$\theta_{T2} = 0.001 \tag{2}$$

in the solutions of the $3 \times 2$ original-type model of BC where all three factors are mobile (see Nakada (2017)), we can expect to derive the basic relationships similar to those in the asymmetrical $3 \times 2$ model. $\theta_{T2}$ is the distributive share of land in sector 2. For the definition of $\theta_{T2}$, see (3).

After Feeny, Williamson (2002, pp. 67-70) applied the simplest type of $3 \times 2$ model, known as the specific factors model, to the nine countries in the preindustrial Third World, including Thailand, using data prior to 1940 (1870 to 1939). He mainly focused on how terms of trade affected relative factor price.

The following questions arise.

(i) What results can we derive if we analyze the $3 \times 2$ model where all three factors are mobile properly?
(ii) What may we conclude if we apply these results to Thailand for the period 1920-1927?

Hardly any study has systematically analyzed question (i), which relates specifically to the sufficient condition for each Rybczynski sign pattern ( $sign[X_j*/V_i*]$ ) to hold in the $3 \times 2$ model, which expresses the factor endowment–commodity output relationships. Nakada (2017) derived the condition. Notably, Nakada (2017) defined the EWS-ratio vector based on the "economy-wide substitution" (hereinafter EWS) originally defined by Jones and Easton (1983) (hereinafter JE) and used it for the analysis. Nakada (2017) concluded that the position of the EWS-ratio vector determines the Rybczynski sign pattern. The author derived a sufficient condition for a strong Rybczynski result to hold (or not to hold). The sufficient condition is that the EWS-ratio vector exists in quadrant IV, in other words, "extreme factors are economy-wide complements."

Thereafter, Nakada (2018) showed that the EWS-ratio vector exists on the line segment.

Using this relationship, he developed a method to estimate the position of the EWS-ratio vector. Nakada (2018) derived a sufficient condition for the EWS-ratio vector to exist in quadrant IV.[4]

According to Suzuki (1983, p. 141), BC contended in Theorem 6 (p. 34) that "if commodity 1 is relatively capital intensive and commodity 2 is relatively labor intensive, an increase in the supply of labor increases the output of commodity 2 and reduces the output of commodity 1. [Moreover, an increase in the supply of capital increases the output of commodity 1 and reduces the output of commodity 2.]" This, in other words, is the implication of "a strong Rybczynski result."

Further, hardly any studies has attempted answering question (ii).[5] That is, hardly any studies has applied the results of Nakada (2017, 2018). Hence, the purposes of this paper are: (i) to apply these results to data from Thailand, and in doing so, to derive the Rybczynski sign patterns for Thailand during the period 1920-1927;[6] (ii) furthermore, based on the results, to investigate whether the labor stock affected the growth of rice output relative to textile output positively, hence, affected the share of exportable sector in national income positively. I restrict the analysis to this period on account of data availability. I have another reason to analyze this period. It appears that both kilograms of white shirting imported per *picul*[7] of rice and rice output increased during that period. It seems appropriate to use that period as the period under study, assuming terms of trade were the main factor to affect rice output.[8]

The value of white shirting imported was larger than that of grey shirting imported during 1923-36, according to Thailand, Statistical Year Book of the Kingdom of Siam (hereinafter, SYB), No. 13, 14, 16, and 18. Therefore, in this article, we use kilograms of white shirting per *picul* of rice as terms of trade.

In addition, the Stolper-Samuelson sign pattern, which expresses the commodity price–factor price relationship, is a dual counterpart in the Rybczynski sign pattern.[9] Regarding this duality, see JE (p. 67) and BC (p. 36, eqs. (31)-(33)). Note, however, that I only analyze the latter.

---

[4] Nakada (2018) analyzed within the framework of the general equilibrium model. For example, Thompson (1995) assumed that production function was of a trans-log type in the US economy. He derived the parameters and Allen's partial elasticities of substitution, using econometrics, and substituted them into the $3 \times 2$ model. Specifically, he substituted the values derived from the partial equilibrium analysis. This seems inconsistent with general equilibrium model.

[5] Nakada (2016) has attempted answering question (ii), referring to the earlier versions of Nakada (2017) and Nakada (2018). Nakada (2016) used kilograms of grey shirting per *picul* of rice as terms of trade, when he analyzed Thailand (1920-1929). During that period, it increased, while rice output decreased. It seems inappropriate to use that period as the period under study, if the terms of trade were the main factor to affect rice output. On this, see Fig.3 and Fig.4. See also Nakada (2015).

[6] According to Ingram (1971, p. 182), import duties were set at 3 percent during the period 1856-1926. Ingram (1971, p. 182) continued, "A new tariff became effective in March 1927…The new duties were still quite low, for the most part. The general rate was increased to 5 percent."

[7] Ingram (1971, p. 40) stated that one *picul* was officially changed from 60.48 kilogram to 60.0 kilogram in 1923, however, the new *picul* was not adopted quickly or uniformly.

[8] For the period 1927-29, both of kilograms of white shirting imported per *picul* of rice and rice output seemed to decrease, hence, it seems plausible to apply the $3 \times 2$ model to this period. However, I am uncertain whether it is plausible to apply it to the period post 1929. Specifically, for the period 1929-1934, kilograms of white shirting imported per *picul* of rice seems to be related negatively to the rice output. On this, see Fig.3 and Fig.4.

I start by deriving the trends of some variables for the period under study. Nakada (2018, p. 14) stated that, in order to estimate the position of the EWS-ratio vector, we need the data about the change in some variables, which requires the data on two time-points to apply his results, whereas normal computable general equilibrium (hereinafter CGE) analysis needs the data for one time-point only in order to estimate the value of basic parameters.

In the model, we consider rice as an exportable (or commodity 1) and cotton textiles as an importable (or commodity 2). We consider land, capital, and labor as the three factors. It seems plausible that cotton products and cotton textiles made in Thailand competed with imported cotton textiles. Feeny (1982, Appendix 2, pp. 162-168) explored the plausibility of the assumptions such as factor mobility, perfect employment, pure competition, and small country assumption, which implies that factor prices and factor endowments are exogenous (see section 2 of this paper). I accept his discussion in this article.

In section 2, I present the theoretical results from Nakada (2017, 2018). In section 3, I conduct analyses for the following.

1. We derive the factor-intensity ranking and show that labor is the middle factor, and land and capital are extreme factors. We assume the factor intensity ranking for middle factor.
2. We derive the trend in the wage for rice and land price for rice.
3. We derive the trend in terms of trade.
4. From the results of 1, 2, and 3, we estimate the factor-price-change ranking, and using Lemma 2, we derive its implication.
5. Using the results of 4, we estimate the sign of aggregate of the rate of change in the input–output coefficient.

In section 4, we conduct analyses for the following.

6. From the results of 3-5, and Theorem 2, we show that the EWS-ratio vector exists in quadrant IV

---

[9] In Section 4, Teramachi (2015, p. 50) showed 12 patterns of '$J$ sign patterns', which express the commodity price–factor price relationships ( $J \equiv \partial \log W / \partial \log P$ , $J = w_i{}^* / p_j{}^*$ in our expression). This is not equivalent to the commodity price–factor price relationships in Nakada (2017) (or $(w_i{}^* - p_j{}^*)/(p_1{}^* - p_2{}^*)$ ). Teramachi did not show a sufficient condition for each $J$ sign pattern to hold systematically. Moreover, Teramachi did not show the process of computation. See Nakada (2017, p. 98, n. 13).

(or subregions P1-P3), in other words, land and capital, extreme factors, are economy-wide complements.

7. From the result of 6 and Theorem 1, we prove that a strong Rybczynski result holds. We derive three Rybczynski sign patterns. However, by making a more detailed estimate, we reduce three candidates to two.

Section 5 concludes the paper. In Appendix A, we derive the equation for the change in the share of exportable sector in national income. In Appendix B, we compute the percentage of net arrivals of Chinese in the population growth in Thailand.

2. Assumptions of the model and some results

Like BC (pp. 22-23), we make the following assumptions. Products and factors markets are perfectly competitive. Supply of all factors is perfectly inelastic. Production functions are homogeneous of degree one and strictly quasi-concave. All factors are not specific and perfectly mobile between sectors, and factor prices are perfectly flexible. The last two assumptions ensure full employment of all resources. The country is small and faces exogenously given world prices; the movement in the price of a commodity is exogenously determined. The movements in factor endowments are also exogenously determined.

For additional definitions of the symbols used and derivations of the basic equations, see Nakada (2017, 2018).

2.1. Factor intensity ranking

Nakada (2017, 2018) have assumed

$$\theta_{T1}/\theta_{T2} > \theta_{L1}/\theta_{L2} > \theta_{K1}/\theta_{K2}, \tag{3}$$

$$\theta_{L1} > \theta_{L2}. \tag{4}$$

where $\theta_{ij}$ is the distributive share of factor $i$ in sector $j$ (that is, $\theta_{ij} = a_{ij}w_i/p_j$). $a_{ij}$ denotes the requirement of input $i$ per unit of output of good $j$ (or the input–output coefficient), $w_i$ is the reward of factor $i$, and $p_j$ is the price of good $j$. Note that $\Sigma_i \theta_{ij} = 1$.

(3) is referred to as the "factor intensity ranking" (see JE (p. 69), BC (pp. 26-27), and Suzuki (1983, p. 142)). This implies that sector 1 is relatively land intensive, sector 2 is relatively capital intensive, labor is the middle factor, and land and capital are extreme factors (Ruffin, 1981, p. 180). JE (p. 70) called (4) the "factor intensity ranking for middle factor." It implies that the middle factor is used relatively intensively in the first industry.

Using these assumptions, we derive the following results.

2.2. Results from Nakada (2017)

In this subsection, we refer to Nakada (2017). The Rybczynski matrix $[X_j{}^*/V_i{}^*]$ (to use Thompson's (1985, p. 619) terminology) in elasticity terms is

$$[X_j{}^*/V_i{}^*] = \begin{bmatrix} X_1{}^*/V_T{}^* & X_1{}^*/V_K{}^* & X_1{}^*/V_L{}^* \\ X_2{}^*/V_T{}^* & X_2{}^*/V_K{}^* & X_2{}^*/V_L{}^* \end{bmatrix}, \qquad (5)$$

For the definitions of the symbols, see (1). The following result has been established already (see (Nakada (2017, Theorem 1)). We have rearranged it below.

**Theorem** 1. We assume the factor intensity ranking as follows.[10]

$$\theta_{T1}/\theta_{T2} > \theta_{L1}/\theta_{L2} > \theta_{K1}/\theta_{K2}, \qquad (6)$$
$$\theta_{L1} > \theta_{L2}. \qquad (7)$$

Further, if the EWS-ratio vector $(S', U')$ exists in quadrant IV (or subregions P1-P3), in other words, if capital and land, extreme factors, are economy-wide complements, a "strong Rybczynski result" necessarily holds. In this case, the Rybczynski sign patterns, as per Thompson's (1985, p. 619) terminology, for subregions P1-P3 are, respectively:

$$\text{sign}[X_j{}^*/V_i{}^*] = \overset{P1}{\begin{bmatrix} + & - & - \\ - & + & + \end{bmatrix}} \overset{P2}{\begin{bmatrix} + & - & + \\ - & + & + \end{bmatrix}} \overset{P3}{\begin{bmatrix} + & - & + \\ - & + & - \end{bmatrix}}. \qquad (8)$$

About subregions P1-P3, see Fig. 1 in Nakada (2017). Each sign pattern expresses the factor endowment–commodity output relationship. For example, the sign of Column 3 shows the labor endowment–commodity output relationship. (8) implies that an increase in the supply of land increases the output of commodity 1 and reduces the output of commodity 2. Moreover, an increase in the supply of capital reduces the output of commodity 1 and increases that of commodity 2. However, it is indeterminate how an increase in the supply of labor affects the outputs of commodities 1 and 2. Three patterns are possible.

The symbols are defined as follows:

---

[10] Assuming $\theta_{L1} < \theta_{L2}$, one can easily show that Theorem 1 holds. One can also show that both of Lemma 2 and Theorem 2 shown below hold.

$$(S', U') = (S/T, U/T) = (g_{LK}/g_{LT}, g_{KT}/g_{LT}), \qquad (9)$$

$$(S, T, U) = (g_{LK}, g_{LT}, g_{KT}), \qquad (10)$$

$$g_{ih} = \Sigma_j \lambda_{ij} \varepsilon^{ij}_h, i, h = T, K, L, \text{ and} \qquad (11)$$

$$\varepsilon^{ij}_h = \partial \log a_{ij} / \partial \log w_h = \theta_{hj} \sigma^{ij}_h, i, h = T, K, L, j = 1, 2 \qquad (12)$$

We call $(S', U')$ the economy-wide substitution (EWS) ratio vector. $S'$ denotes the relative magnitude of EWS between factors $L$ and $K$ compared to EWS between factors $L$ and $T$. $U'$ denotes the relative magnitude of EWS between factors $K$ and $T$ compared to EWS between factors $L$ and $T$. $g_{ih}$ is the EWS between factors $i$ and $h$, as defined by JE (p. 75). $g_{ih}$ is the aggregate of $\varepsilon^{ij}_h$. JE (p. 75) stated, "Clearly, the substitution terms in the two industries are always averaged together. With this in mind we define the term $\sigma^i_k$ to denote the economy-wide substitution towards or away from the use of factor $i$ when the $k$th factor becomes more expensive, under the assumption that each industry's output is kept constant.…" $\sigma^{ij}_h$ is Allen's partial elasticity of substitution between the $i$th and the $h$th factors in the $j$th industry. For additional details about these symbols, see BC (p. 24) and Sato and Koizumi (1973, pp. 47-49). $\lambda_{ij}$ is the proportion of the total supply of factor $i$ in sector $j$ (that is, $\lambda_{ij} = a_{ij} X_j / V_i$). Note that $\Sigma_j \lambda_{ij} = 1$. $X_j$ denotes the amount of good $j$ produces $(J = 1, 2)$. For the definition of $a_{ij}$ and $V_i$, see (3) and (1), respectively.

We obtain (see JE (p.72, n.9))

$$\lambda_{ij} = (\theta_j / \theta_i) \theta_{ij}, \qquad (13)$$

where $\theta_j$ and $\theta_i$ denote, respectively, the share of good j and factor i in total income. That is, $\theta_j = p_j X_j / I$, $\theta_i = w_i V_i / I$, where $I = \Sigma_j p_j X_j = \Sigma_i w_i V_i$. Note that $\Sigma_j \theta_j = 1$, $\Sigma_i \theta_i = 1$. See BC (p.25, eq. (16)).

We may also define the following ($i \neq h$) (Nakada, 2017, eq. (45)):

Factors $i$ and $h$ are economy-wide substitutes if $g_{ih} > 0$, and
Factors $i$ and $h$ are economy-wide complements if $g_{ih} < 0$. (14)

2.3. Building on the results of Nakada (2018)

In this subsection, we refer to Nakada (2018). Note that we add some original equations. For ease of notation, we define (Nakada, 2018, eq. (34))

$$(X, Y, Z) = (w_T{}^* - p_1{}^*, w_K{}^* - p_1{}^*, w_L{}^* - p_1{}^*), \tag{15}$$

Next, we define $P$:

$$P = (p_1/p_2)^* = p_1{}^* - p_2{}^*. \tag{16}$$

$P$ is the rate of change in terms of trade.

The following result has been established (Nakada, 2018, Lemma 2).[11]

Lemma 2 We assume the factor intensity ranking and the change in the relative price of goods as follows.

$$\theta_{T1}/\theta_{T2} > \theta_{L1}/\theta_{L2} > \theta_{K1}/\theta_{K2}, \quad \theta_{L1} > \theta_{L2}, \tag{17}$$
$$P = p_1{}^* - p_2{}^* > 0. \tag{18}$$

And, further, if we assume the factor-price-change ranking as follows (from Lemma 1, this assumption is plausible enough)

$$X > Z > Y \leftrightarrow w_T{}^* > w_L{}^* > w_K{}^*, \tag{19}$$

the signs A, B, C, D are possible. That is,

$$\begin{array}{cccc} & A & B & C & D \\ (a_{T0}{}', a_{K0}{}', a_{L0}{}') = & (-,+,-), (-,+,+), (+,+,-), (-,-,+), & & & \end{array} \tag{20}$$
$$(a_{Tj}{}^*, a_{Kj}{}^*, a_{Lj}{}^*) = (-,+,-), (-,+,+), (+,+,-), (-,-,+), \ j = 1.2, \tag{21}$$

where $a_{i0}{}' = \sum_j \lambda_{ij} a_{ij}{}^*, i = T, K, L$. $a_{i0}{}'$ is the aggregate of $a_{ij}{}^*$ (or the rate of change in the input-output coefficient).

Equations (20) and (21) imply that signs of $a_{i0}{}'$ and $a_{ij}{}^*$ are not so arbitrary. The following result has been established (Nakada, 2018, Theorem 1).

---

[11] I omit Lemma 1 in Nakada (2018).

**Theorem** 2. We assume the factor intensity ranking and the change in the relative price of goods as follows.

$$\theta_{T1}/\theta_{T2} > \theta_{L1}/\theta_{L2} > \theta_{K1}/\theta_{K2}, \theta_{L1} > \theta_{L2}, \quad (22)$$
$$P = p_1{}^* - p_2{}^* > 0. \quad (23)$$

The EWS-ratio vector $(S',U')$ exists on the EWS-ratio vector line segment (or line segment AB). Using this relationship, we can estimate the position of the EWS-ratio vector. For example, if we assume (from Lemma 2, these assumptions are plausible enough)

$$X > Z > Y \leftrightarrow w_T{}^* > w_L{}^* > w_K{}^*, \quad (24)$$
$$(a_{T0}', a_{K0}', a_{L0}') = (+,+,-), \quad (25)$$

the Cartesian coordinates of intersection points A and B are, respectively,

$$\left(\frac{-W_{TL}}{W_{KL}}, \frac{\theta_L}{\theta_K}\frac{-W_{LT}}{W_{KT}}\right) = \left(\frac{-(w_T{}^*-w_L{}^*)}{(w_K{}^*-w_L{}^*)}, \frac{\theta_L}{\theta_K}\frac{-(w_L{}^*-w_T{}^*)}{(w_K{}^*-w_T{}^*)}\right) = (+,-), \quad (26)$$

$$\left(\frac{a_{K0}'}{a_{T0}'}\theta_{KT}, \frac{a_{K0}'}{a_{L0}'}\right) = (+,-). \quad (27)$$

Hence, both of points *A* and *B* are in quadrant IV, and, point *A* is on the left-hand side of point *B*. The line segment *AB* exists in quadrant IV. Hence, the EWS-ratio vector is in quadrant IV and satisfies

$$0 < \frac{-W_{TL}}{W_{KL}} < S' < \frac{a_{K0}'}{a_{T0}'}\theta_{KT}, 0 > \frac{\theta_L}{\theta_K}\frac{-W_{LT}}{W_{KT}} > U' > \frac{a_{K0}'}{a_{L0}'}. \quad (28)$$

In this case, capital and land, extreme factors, are economy-wide complements. Hence, a strong Rybczynski result holds, that is, three of the Rybczynski sign patterns hold (see Theorem A.1).

About Points A and B, see **Fig.** 1 in Nakada (2018). In sum, (23)-(25) imply the following. Terms of trade increase. The rate of change in real reward for labor is intermediate (or moderate), and the rates of change in real reward for land and capital are extreme. The aggregate of $a_{Tj}{}^*$ and $a_{Kj}{}^*$ (or the rates of change in the input–output coefficients of land and capital, respectively) increase, but the aggregate of $a_{Lj}{}^*$ (or the rate of change in the input–output coefficient of labor) decreases.

The symbols are defined as follows:

$$W_{ih} = w_i{}^* - w_h{}^* = (w_i / w_h)^*, i, h = T, K, L, i \neq h, \quad (29)$$

$$\theta_{ih} = \theta_i / \theta_h, i, h = T, K, L, i \neq h, \quad (30)$$

According to Nakada (2018, eq. (C6)), we can compare the following equations:

$$P > 0, X > Z > -P(>[-\theta_{T1}/(\theta_{T1} - \theta_{T2})]P) , \quad (31)$$

$$P > 0, X > Z(>-P) > Y , \quad (32)$$

Equations (31) and (32) are sometimes useful when we apply. ( $[-\theta_{T1}/(\theta_{T1}-\theta_{T2})]P$ , $[-\theta_{T1}/(\theta_{T1}-\theta_{T2})]P$ ) denote the Cartesian coordinates of the intersection of Lines Y and Z. If (31) holds, we derive (32), but not vice versa.

3. Proving the assumptions in Theorem 2 to hold

We use some of the derived results from the $3 \times 2$ model for the period 1920-1927.

First, we derive the factor intensity ranking. Next, we prove whether (23)-(25) hold for the period 1920-1927. We can easily show that (23) holds. We prove whether (24) and (25) hold.

3.1. Factor intensity ranking

We estimate $\lambda_{i1}$, $i = T, K, L$. Here, recall (11), that is, $\lambda_{ij}$ is the proportion of the total supply of factor $i$ in sector $j$ (that is, $\lambda_{ij} = a_{ij}X_j/V_i$). Note that $\Sigma_j \lambda_{ij} = 1$.

(i) $\lambda_{T1}$: Using Thailand, SYB, No. 18, Ingram (1971), and Yamamoto (1998), **Table** 1 presents the total area planted with 11 crops in Thailand, that consist of principal crops in 1927-28 and three other crops in 1924 and 1937. Thailand's principal crops include rice, tobacco, maize, cotton, peas, sesame, pepper, and coconut. Three other crops include fruits, vegetables, and sugarcane. We classify five crops (rice, peas, sesame, pepper, and coconut) as exportables.[12] In 1927-1928, we estimate that the area planted with exportables (18,672,247 *rai*[13]) comprised 89.5% of the total area (20,861,096 *rai*). We consider the percentage of area planted with exportables as $\lambda_{T1}$.

*Insert Table 1 here*

(ii) $\lambda_{L1}$: In 1929, the labor engaged in agriculture (6,245,358 persons) comprised about 83.1% of the total labor force (7,519,757 persons) in Thailand (SYB, No. 18, p. 88). We consider the percentage of labor engaged in agriculture as $\lambda_{L1}$.

---

[12] According to Thailand, SYB, No. 13, Thailand exported beans and peas, pepper, and copra during 1923-27. We estimate that sesame was also exported.

[13] 6.25 *rai* was equal to one hectare.

(iii) $\lambda_{K1}$: There are no available data pertaining to the amount of capital invested in Thailand's rice sector during the period under study. It seems that the sector did not require much capital. Buffaloes and ploughs appear to have been essential farm implements (Feeny, 1982, pp. 40-41, Table 4-2). Hence, it seems plausible that $\lambda_{L1} > \lambda_{K1}$. Therefore,

$$\lambda_{T1} > \lambda_{L1} > \lambda_{K1} \tag{33}$$

is plausible. Accordingly,

$$(\lambda_{T1}, \lambda_{K1}, \lambda_{L1}) = (0.90, ?, 0.83), \quad (\lambda_{T2}, \lambda_{K2}, \lambda_{L2}) = (0.10, ?, 0.17). \tag{34}$$

Feeny assumed $\lambda_{T1} = 1$ and estimated that $\lambda_{L1} > \lambda_{K1}$ (Feeny, p. 27, Appendix 2, pp. 169-170). We use the same estimation.

We can easily show that

$$\lambda_{T1} > \lambda_{L1} > \lambda_{K1} \leftrightarrow \theta_{T1}/\theta_{T2} > \theta_{L1}/\theta_{L2} > \theta_{K1}/\theta_{K2}. \tag{35}$$

Recall (3), that is, factor intensity ranking. This implies that sector 1 is relatively land intensive, sector 2 is relatively capital intensive, labor is the middle factor, and land and capital are extreme factors.

For example, Kamol Odd computed the production cost of rice for a sample of 106 households in Ban Chan village in Central Thailand in 1948 (Kamol, 1955, p. 279, Table 38). He derived the shares of three factors, namely land, capital, and labor, as 22.4%, 26.3%, and 51.3% respectively. This implies that

$$(\theta_{T1}, \theta_{K1}, \theta_{L1}) = (0.22, 0.27, 0.51). \tag{36}$$

The sample households planted 3,528 *rai* in total, which included transplanted rice of 3,030 *rai* (86%) and broadcasted rice of 498 *rai* (14%) (Kamol, 1955, p. 213). There was almost no floating rice (Kamol, 1955, p. 212). Only one household broadcasted it (Kamol, 1955, p. 105).

These data were surveyed in 1948. Feeny (1982, p. 27) also referred to Kamol Odd (1955).

The production cost of cotton is not available. Hence, the data for $(\theta_{T2}, \theta_{K2}, \theta_{L2})$ are not available, and it is not possible to determine which of the following equations holds:

$$\theta_{L1} > \theta_{L2} \text{ or} \tag{37}$$
$$\theta_{L1} < \theta_{L2}. \tag{38}$$

Recall (4). This is the factor intensity ranking for the middle factor. For example, if (37) holds, the middle factor is used relatively intensively in sector 1. We assume (37) holds. Nakada (2017, 2018) only assume that (35) and (37) hold. Neither assume that (38) holds.

3.2. Factor price

We analyze the real wage for the period 1864-1938. Some authors have referred to the wage in Thailand before World War II (see Skinner (1957, p. 174), Ingram (1964, pp. 113-117), Feeny (1982, p. 34), Sompop (1989, pp. 164-166, Table 6.4, p. 168, Table 6.7), and Porphant (1998, pp. 81-85)).

Ingram (1964, p. 112) stated, "The trend of this rice wage-rate...was downward from the 1820s to about 1910, after which it recovered slightly in the 1920s and rose sharply with the onset of the depression in 1930." Further, Feeny (1982, p. 29-34) wrote, "For urban workers, the real wage in cloth increased while the real wage in rice decreased over the whole period [around 1864-1901]…In the period from 1901-1921 urban real wages, in contrast to rising rural incomes, decreased…From 1921 to 1938 the situation reversed itself; urban real wages were clearly increasing." However, these statements do not provide exact values.

**Fig.** 1 plots the daily wages of unskilled (*coolie*) laborers in *picul* of rice in Bangkok for the period 1864-1938.[14] The wage data are not available to the extent required. Ingram (1964, p. 112) noted, "The sharp drop [of the rice wage-rate] in 1919 was the result of a severe crop failure in which rice prices rose drastically and an embargo was put on rice."

*Insert Fig. 1 here*

Using this information, we can trace the trends in the rice wage-rate as follows.

(i)     During 1889-1920, it decreased.
(ii)    During 1920-1927, it decreased again.
(iii)   During 1927-1934, it increased significantly.
(iv)    During 1934-1938, it decreased.

Next, we analyze the rent in the period 1880-1941. As data on rent are not available to the extent needed, we attempt to use the land price instead. The lack of land prices in the same area leads me to use the data provided in Johnston (1975) and Thailand, SYB, Nos. 18 and 19.

---

[14] Specifically, we referred to the daily wage of unskilled laborers in 1914 from Thailand, SYB, No. 4, which Ingram (1964) did not show. It was 0.75 baht as in 1915 and 1916.

Johnston (1975, p. 121) presented the land prices on the east bank of Chaophraya River for the period 1880-1904. The prices were originally derived from a variety of documents from Thailand and Van der Heide (1903). Johnston stated, "Data is extremely limited…It confirms the impression of observers at the time that land prices, especially on the east bank, were increasing rapidly at the turn of the century." Thailand, SYB, No. 18 and 19 list the nominal value of paddy land mortgaged per *rai* from 1915-1941. On these data, see also Feeny (1982, p. 137, Tables A1-8).

Using the data of Johnston (1975, p. 121) and Thailand, SYB, No. 18 and 19, **Fig.** 2 presents the land price in terms of rice for the periods 1880-1904 and 1915-1941. Unfortunately, land price data for the period 1904-1915 are not available to the extent needed. Hence, we estimate them. For example, while Feeny listed land prices in specific areas (Feeny 1982, p. 135, Tables A1-7), some data are still missing.

*Insert Fig. 2 here*

The following trend may be observed with respect to the real land price measured by rice.

(i) During 1880-1904, the real land price measured by rice increased.
(ii) During 1904-1920, we estimate that it decreased.
(iii) During 1920-1931, it increased.
(iv) During 1931-1938, it decreased.

3.3. Terms of trade

We analyze the terms of trade, that is, kilograms of grey (and white) shirting imported per *picul* of rice. **Fig.** 3 presents the terms of trade for the period 1864-1945.

The following trend for terms of trade (kilograms of white shirting per *picul* of rice) is evident.

*Insert Fig. 3 here*

1. During 1864-1898, terms of trade increased.
2. During 1898-1920, terms of trade decreased.
3. During 1920-1927, terms of trade increased.
4. During 1927-1934, terms of trade decreased.[15]

---

[15] Suehiro (2000, p. 74-75) noted that kilograms of white shirting per *picul* of rice slumped during the

5. During 1934-1941, terms of trade increased.

3.4. Factor-price-change ranking

We compare the trends of wage in rice, land price in rice, and terms of trade in the period 1864-1945. The results indicate that the land price in rice changed with some time lag. For example, around 1920-1927, the terms of trade and land price in rice increased, while the real wage measured by rice decreased.

From the data, we derive a rate of change for these variables for the period 1920-1927:[16]

$$P = p_1{}^* - p_2{}^* = +176.6\% > 0, \quad X = w_T{}^* - p_1{}^* = +22.1\% > 0, \quad Z = w_L{}^* - p_1{}^* = -12.5\% < 0 \tag{39}$$

where P is the rate of change in the kilograms of white shirting per picul of rice. Hence, we show that (see (31))

$$P > 0, \quad X > Z > -P. \tag{40}$$

From (40), we derive (see (32))

$$P > 0, \quad X > Z > Y. \tag{41}$$

Hence, we have shown that (23) and (24) hold. However, (25) is not self-evident.

Here, recall Lemma 2. If (41) holds, (18) and (19) hold; hence, (20) holds. That is,

$$\begin{array}{cccc} A & B & C & D \end{array}$$

$$(a_{T0}{}', a_{K0}{}', a_{L0}{}') = (-,+,-), (-,+,+), (+,+,-), (-,-,+). \tag{42}$$

We can show which signs of A, B, C, and D are possible for sector *j*.

3.5. Estimating the sign of $a_{i0}{}'$ (aggregate of the rate of change in the input–output coefficient)

We analyze the sign of $a_{i0}{}'$ (aggregate of the rate of change in the input–output coefficient).

We estimate the sign of $a_{T1}{}^*$, that is, the rate of change in the input–output coefficient of land in sector 1. Multiplying the change in the average yield of rice by –1 helps us observe the change in the input–output coefficient of sector 1.

---

period 1926-1934.
[16] We do not need to show the rate of change per year.

**Fig.** 4 illustrates the production, area sown, and average yield of rice in Thailand during 1918-1936. The 3-year moving average of the average yield is also depicted. Feeny stated, "While it is recognized that the data are not absolutely reliable and that under-reporting of area and output was probably prevalent, it is argued…that the trends in the series are probably reliable" (Feeny, 1982, p. 48).

*Insert Fig. 4 here*

Note that the year 1919 experienced a severe loss of rice crop as mentioned earlier (see Ingram, 1964, p. 112). Furthermore, Kaida (1978, p. 208) referred to information from the Royal Irrigation Department (RID), Thailand and noted the various losses in the country's rice crop during that period. The paddy area around the Chaophraya River Basin recorded extensive damage. In 1917, 21.0% of the area witnessed flooding. In 1919, there was a very severe drought, affecting 43.4% of the area. The year 1929 recorded a moderate drought affecting 19.5% of the area. Severe drought and severe flooding occurred in 1939 and 1942, damaging 31.7% and 34.3% of the area, respectively.

Thus, the trend in the average yield of rice (kg/*rai*) may be determined as follows.

(i)     The average yield (kg/*rai*) decreased for the period 1920-1927.[17]
(ii)    It decreased again for the period 1927-1929.
(iii)   It increased for the period 1929-1932.
(iv)   It decreased for the period 1932-1936.

Hence, we derive

$$a_{T1}^* = (+) \text{ for 1920-1927,} \tag{43}$$

$$a_{T1}^* = (+) \text{ for 1927-1929,} \tag{44}$$

---

[17] Langmoya (1978) studied the reign of King Chulalongkorn, the fifth monarch of Siam (1868-1910). Langmoya (p. 230) stated that about 50 % of paddy field was broadcasted in the Central Plain., referring to Ministry of Agriculture (1961), *Agriculture in Thailand*, p. 6. I was unable to identify the time period Langmoya's statement is attempting to support. Feeny (1982, p. 44) stated, referring to Indra Montri (1930, p. 8), "In 1930 roughly 30 per cent of the paddy output in the Central Plain was accounted for by transplanted rice and 70 per cent by broadcast." The diffusion of broadcast rice must have contributed to the decrease of the average yield in the Central Plain during the period 1920-27. On the other hand, Sompop (1989, p. 83) stated, referring to Langmoya (1978, p. 230-233) and Intaramontri (1930) cited in Feeny (1982, p. 44-45), "broadcast rice increased about half to three quarters of total areas in the Central Plain by the late 1900's to the 1930's." See also Sompop (1989, p. 68, 170).

$$a_{T1}^* = (-) \text{ for 1929-1932, and} \tag{45}$$
$$a_{T1}^* = (+) \text{ for 1932-1936.} \tag{46}$$

We estimate the sign of $a_{T2}^*$, that is, the rate of change in the input–output coefficient of land in sector 2. Multiplying the change in the average yield of cotton by –1 helps us observe the change in the input–output coefficient of sector 2.

**Fig.** 5 shows the production, area sown, and average yield (kg/*rai*) of cotton in Thailand during 1918-1936. We also indicate the 3-year moving average of the average yield. It appears that the official data pertaining to the area sown are under-reported (see, e.g., Sugawara (2000, p. 89)). The yields in 1918, 1929, 1931, and 1935 were quite high. We assume that the trends in the series are probably reliable.

*Insert Fig. 5 here*

Based on this information, we can decipher the trend in the average yield of cotton as follows.

(i)   The average yield decreased for the period 1920-1927.
(ii)  It increased for the period 1927-1936.

Hence, we derive

$$a_{T2}^* = (+) \text{ for 1920-1927 and} \tag{47}$$
$$a_{T2}^* = (-) \text{ for 1927-1936.} \tag{48}$$

From (43) and (47), we derive the following for the period 1920-1927:

$$a_{T0}' = (+). \tag{49}$$

From (49) and (42), we derive

$$(a_{T0}', a_{K0}', a_{L0}') = (+, +, -). \tag{50}$$

We show that (25) holds.

4. Deriving the Rybczynski sign patterns

In this section, we derive the Rybczynski sign patterns.

4.1. Rough estimate

In sum, we derived (35), and assumed (37). Using (41) and (50), we derive the following for the period 1920-1927:

$$P > 0, X > Z > Y \leftrightarrow w_T^* > w_L^* > w_K^*, (a_{T0}', a_{K0}', a_{L0}') = (+, +, -). \quad (51)$$

(51) is equivalent to (23)-(25) in Theorem 2. This implies that the EWS ratio vector (S', U') exists in quadrant IV.

Hence, from Theorem 1, we determine the Rybczynski sign patterns for each subregion as seen below (see (8)).

$$\text{sign}\left[X_j^*/V_i^*\right] = \overset{P1}{\begin{bmatrix} + & - & - \\ - & + & + \end{bmatrix}} \overset{P2}{\begin{bmatrix} + & - & + \\ - & + & + \end{bmatrix}} \overset{P3}{\begin{bmatrix} + & - & + \\ - & + & - \end{bmatrix}}. \quad (52)$$

Each sign pattern shows the factor endowment–commodity output relationship. Notably, the sign in Column 3 shows the labor endowment–commodity output relationship.

Therefore, we can make the following statements.

(i) If the EWS ratio vector (S', U') exists in subregion P1, the effects of labor endowment on commodity output in sector 1 and sector 2 are negative and positive, respectively.

(ii) If the EWS ratio vector exists in subregion P2, the effects of labor endowment on commodity output in both sectors 1 and 2 are positive.

(iii) If the EWS ratio vector exists in subregion P3, the effects of labor endowment on commodity output in sector 1 and sector 2 are positive and negative, respectively.

From (52) and (5), we derive the following for P1, P2, and P3, respectively:

$$X_1^*/V_L^* - X_2^*/V_L^* = (-) - (+) = (-), \quad (53)$$

$$X_1^*/V_L^* - X_2^*/V_L^* = (+) - (+) = (?) \text{ and} \quad (54)$$

$$X_1^*/V_L^* - X_2^*/V_L^* = (+) - (-) = (+). \quad (55)$$

Recall (1), that is,

$$(X_1/X_2)^*/V_L^* = X_1^*/V_L^* - X_2^*/V_L^* = (+).  \qquad (1)$$

The sign of the left-hand side of (1) shows how labor endowment affects the commodity output in sector 1 relative to commodity output in sector 2. (53) belies Feeny's (1982, p. 28) statement. (54) might be contrary to it, while (55) is not against it. At the very least, Feeny's statement that the growth in the labor stock was responsible for Thailand's increased rice output relative to textile output is not self-evident.

4.2 More detailed estimate

We can make a more detailed estimate. We reduce three candidates to two. From (39), we derive

$$w_L^* - p_1^* = -12.5\% < 0, \qquad (56)$$
$$w_L^* - p_2^* = w_L^* - p_1^* + (p_1^* - p_2^*) = -12.5\% + 176.6\% = 164.1\% > 0. \qquad (57)$$

On the other hand, Nakada (2018, Corollary 1) derived the result as the following.

If the equation shown below holds,

$$\frac{\theta_{K1}}{\theta_{T1}} < \frac{-W_{TL}}{W_{KL}} < S' < \frac{a_{K0}'}{a_{T0}'}\theta_{KT} < \frac{\theta_{K2}}{\theta_{T2}}, \qquad (58)$$

both Points A and B exist in the subregion P2. Hence, the EWS-ratio vector exists in the subregion P2. The sufficient condition for (58) is the set of equations shown below.

$$\frac{-W_{TL}}{W_{KL}} < \frac{\theta_{K2}}{\theta_{T2}} \leftrightarrow w_L^* - p_2^* = (+) > 0, \qquad (59)$$

$$\frac{\theta_{K1}}{\theta_{T1}} < \frac{-W_{TL}}{W_{KL}} \leftrightarrow w_L^* - p_1^* = (-) < 0 \text{, and} \qquad (60)$$

$$\frac{a_{K0}'}{a_{T0}'}\frac{\theta_K}{\theta_T} < \frac{\theta_{K2}}{\theta_{T2}}. \qquad (61)$$

Apparently, (56) and (57) satisfy (60) and (59), respectively. However, it is uncertain whether the data of Thailand 1920-1927 satisfy (61) because of data availability. Therefore, Point A exists in subregion P2. Point B exists in subregion P2 or P1. Hence, the EWS-ratio vector exists in the subregion P2 or P1. Hence, (53) and (54) are possible, and (55) is impossible. Hence, Feeny's statement shown above is not self-evident.

5. Conclusion

This paper showed that a certain pattern of factor intensity ranking, as shown in (3), holds for Thailand. Moreover, we assume that the factor intensity ranking of the middle factor, as shown in **(4)**, holds. We can draw the following conclusions for the data pertaining to Thailand for the period 1920-1927. The EWS ratio vector (S', U') exists in quadrant IV, therefore, Capital and land, extreme factors, were economy-wide complements.[18] Hence, a "strong Rybczynski result" necessarily holds. We derived three of the Rybczynski sign patterns. However, by making a more detailed estimate, we could reduce three candidates to two. That is, the EWS-ratio vector exists in subregion P2 or P1.

The results imply as follows: Feeny's (1982, p. 28) statement that the growth in labor (or middle factor) stock was responsible for the large growth in rice output relative to textile output in Thailand might not necessarily hold. The labor stock might not affect the share of exportable sector in national income positively. On this, see (A.6).

This paper showed that the percentage of net arrivals of Chinese during 1920-21 to 1926-27 in the population growth during 1920-1927 was not as large as expected (See Appendix B). In summary, even if labor stock had affected the share of exportable sector positively, the effect of Chinese immigration on it would not have been large.

If we wish to derive the sign of the left-hand side of (1) with certainty, we would need to conduct the analysis differently. On this, see Appendix B in Nakada (2016).

We show how factors other than labor stock affected relative output growth during 1920-1927. From (52) and (5), we derive

$$X_1^*/V_T^* - X_2^*/V_T^* = (+) - (-) = (+), \tag{62}$$

$$X_1^*/V_K^* - X_2^*/V_K^* = (-) - (+) = (-). \tag{63}$$

From the above, land (resp. capital) stock affected relative output growth positively (resp. negatively). We can show that terms of trade affect it positively in the $3 \times 2$ model. We do not show the proof.

Feeny (1982, p. 22) stated, referring to Caldwell (1967, p. 32), "For 1900 to 1930…one-third of the population growth was the direct result of [Chinese net] immigration, and with the inclusion of locally born children, as much as 40 percent of the growth in that period would be attributable to [Chinese net] immigration."[19] Feeny (1982) gave the impression that the effect of

---

[18] It is implausible to assume the functional form of, such as Cobb-Douglas, or all-constant CES in each sector, which do not allow any two factors to be Allen-complements. See Nakada (2018).

[19] Caldwell (1967, p. 32) stated, "Between 1900 and 1930 this immigration [or the net intake of Chinese immigration] probably accounted directly for at least one-third of all population growth, and

Chinese net immigration on Thai development, including the change in relative factor prices, was large during that period.[20] We can easily show that this statement of Feeny is implausible, if we compare Table B2 and Table B3. The percentage of net arrivals of Chinese during 1900 to 1929-30 (740 thousand persons) in the population growth during 1900-1930 (5,713 thousand persons) was only 13.0%.

In this article, we have investigated the sign of the effect of exogenous variables on relative output growth.[21] From the Rybczynski sign patterns derived, we can derive the Stolper-Samuelson sign patterns. See Nakada (2017, Theorem 1). Using this result, for example, we can simulate real wage for the period 1920-1927, in order to implement a counterfactual experiment.[22] However, I am afraid that the rate of change in terms of trade derived for 1920-27 is too large to derive a plausible result. It might have been smaller outside Bangkok.

Appendix A: The equation for the change in the share of exportable sector

We derive the equation for the change in the share of exportable sector in national income. For the definitions of the symbols used here, see (13).

$$\theta_j = p_j X_j / I, \; j = 1,2. \tag{A1}$$

$$\theta_1 + \theta_2 = 1. \tag{A2}$$

Totally differentiate eq. (A1) to have:

$$\theta_j^* = p_j^* + X_j^* - I^*, \; j = 1, 2. \tag{A.3}$$

---

if we include locally born children, perhaps two-fifths." Caldwell did not show the references. Later, Carmichael (2008, p. 11) referred to the Caldwell (1967, p. 32)'s statement without any criticism.

[20] In Chapter 3, Feeny (1982, p. 28) concluded, "the general equilibrium model highlights the powerful effects of the terms of trade and the population growth on Thai development, especially the growth in rice output [relative to textile output] and trends in relative factor price." Moreover, in Appendix 2, Feeny (1982, p. 168) stated, "Chapter 3 stressed the importance of the terms of trade and population growth in explaining the events which occurred…much of the growth in population is due to immigration which is at least in part exogenous."

[21] Normally, we do the sensitivity analysis, when we simulate using the CGE model. That is, we confirm whether the results are robust to the choice of the parameter values, such as elasticity parameters. On the robustness of the CGE models, for example, see Barbara (1994). In this article, we have not done a simulation study. Our model is, what you call, the analytical general equilibrium model. At least, we have known that the position of the EWS-ratio vector determines the Rybczynski sign pattern.

[22] On the counterfactual experiment without net immigration, see, for example, Williamson *et al.* (1993) and Hatton and Williamson (1998, p. 212). Theoretically, from Theorem 2, to do a plausible simulation, we should assume that the EWS-ratio vector exists on the EWS-ratio vector line segment.

From eq. (A.3), we have

$$\theta_1^* - \theta_2^* = (p_1^* + X_1^*) - (p_2^* + X_2^*). \tag{A.4}$$

Totally differentiate eq. (A2) to have:

$$\theta_1 \theta_1^* + \theta_2 \theta_2^* = 0. \tag{A.5}$$

Eliminate $\theta_2^*$ from eqs (A.4) and (A.5) to derive:

$$\theta_1^* = \theta_2[(p_1^* - p_2^*) + (X_1^* - X_2^*)]. \tag{A.6}$$

This equation implies that the growth of rice output relative to textile output is related to the growth of the share of exportable sector positively.

Equation (A.5) implies as follows. For example, we assume $\theta_1 = 0.8, \theta_2 = 0.2$. If the share of exportable sector increases 1%, the share of importable sector decreases 4%.

Appendix B: The percentage of net arrivals of Chinese in the population growth in Thailand

In this appendix, we compute the percentage of net arrivals of Chinese in the population growth in Thailand. Of course, the growth of population is different from that of labor stock. However, we do not discuss this further.

**Table** B1 shows the comparison of the estimates of total arrivals and departures of ethnic Chinese, all Thailand for 1918-34. In general, the amount of net arrivals of Chinese computed from Skinner (1957) is larger than that computed from Thailand, SYB, No.18. Specifically in 1929-30, the former is far larger than the latter. **Table** B2 shows the estimated total population in Thailand during 1900 to 1950, referring to Kobayashi (1984) and Bourgeois-Pichat (1960). The population in Kobayashi (1984) is smaller than that in Bourgeois-Pichat (1960) during 1920-33.

These data imply as follows. Specifically, during 1918 to 1931, mass influx of Chinese occurred. During 1920-21 to 1926-27, the total of net arrivals of ethnic Chinese was 236.2 thousand persons, according to Skinner (1957). During 1920-1927, the population growth in Thailand was 1,555 thousand persons, based on Kobayashi (1984). Therefore, the percentage of the former in the latter was 15.2%.

**Table** B3 shows the estimated arrivals and departures of ethnic Chinese in Thailand during 1900 to 1955, referring to Skinner (1957).[23]

---

[23] For the data in **Table** B3, during the period 1900 to 1905 and the period 1940 to 1955, the year

*Insert Table A1, Table A2, and Table A3 here*

---

begins on January 1, however, during the period 1906-07 to 1939-40, it ran from April 1 to March 31. Ingram (1971, p. 4) stated, "Until 1940 the year in Thailand ran from April 1 to March 31… Since 1940 the year begins on January 1…" In this article, the year ran from April 1 to March 31 in general.

Equilibrium. *Journal of Environmental Economics and Management* 6: 187-198.

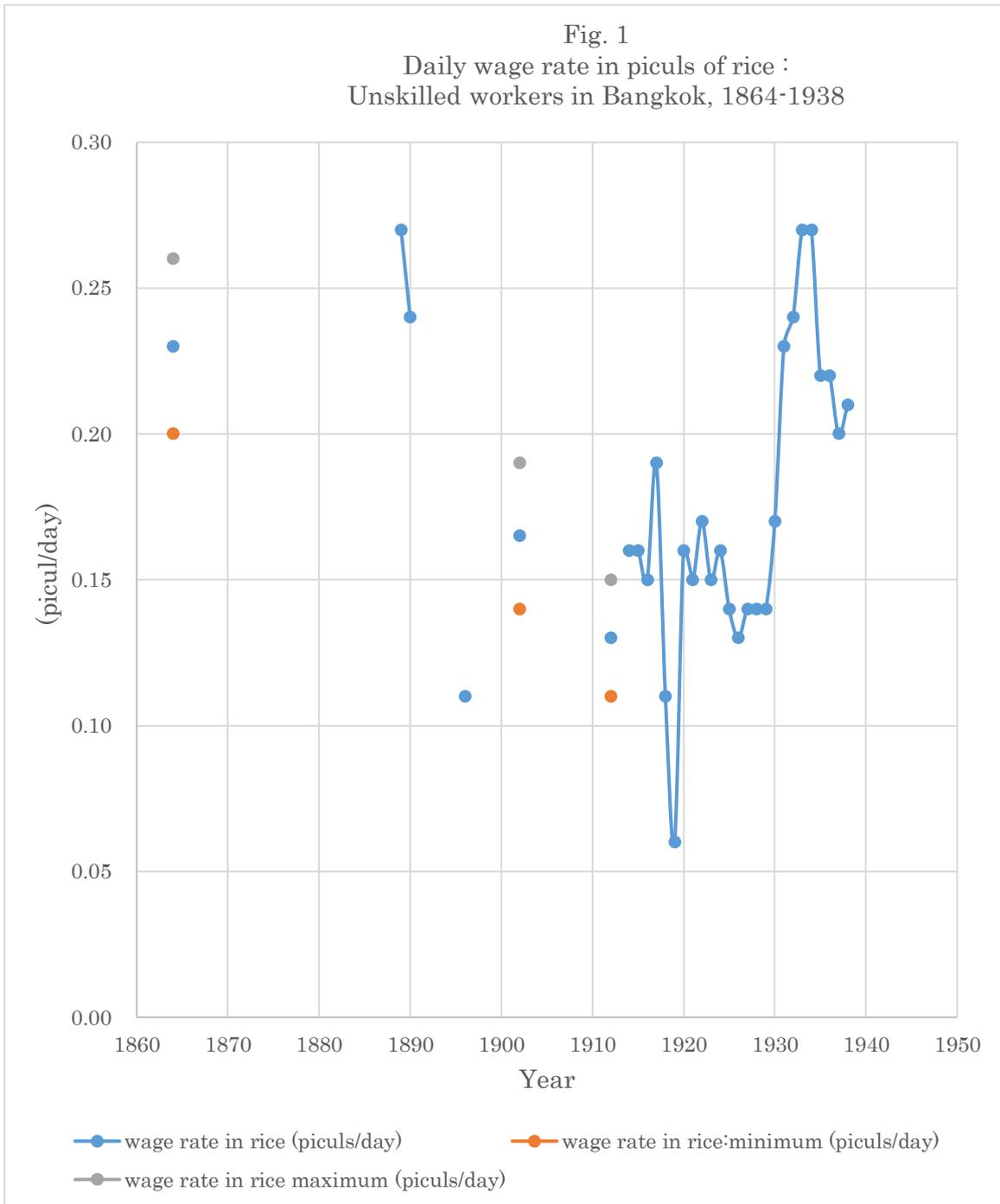

Source: Wage rates: 1864-1912: Ingram (1964, p115, TableIII). 1914-38: SYB, No. 4, 8, 13, 18, and 20.

Price of rice (baht/picul) is computed from Ingram (1964, p120, Appendix A). Note: 1picul = 60.48kg.

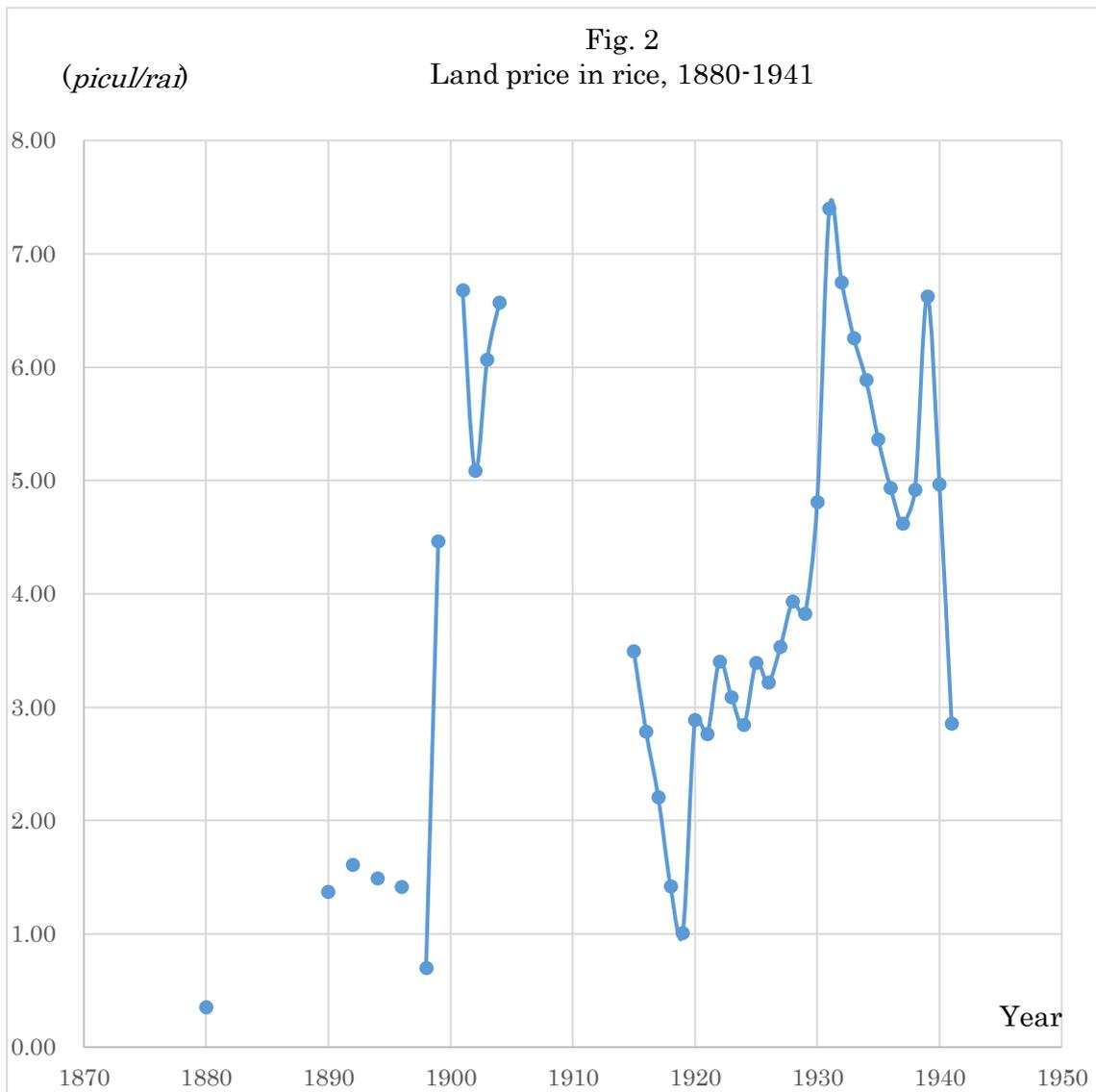

Fig. 2 Land price in rice, 1880-1941

Source: Land prices are from Johnston (1976, Table1, p. 121) and SYB, No.18-19.;
Prices of rice (*baht/picul*) are computed from Ingram (1964, p. 120, Appendix A).

Note: 6.25rai = 1hectare.

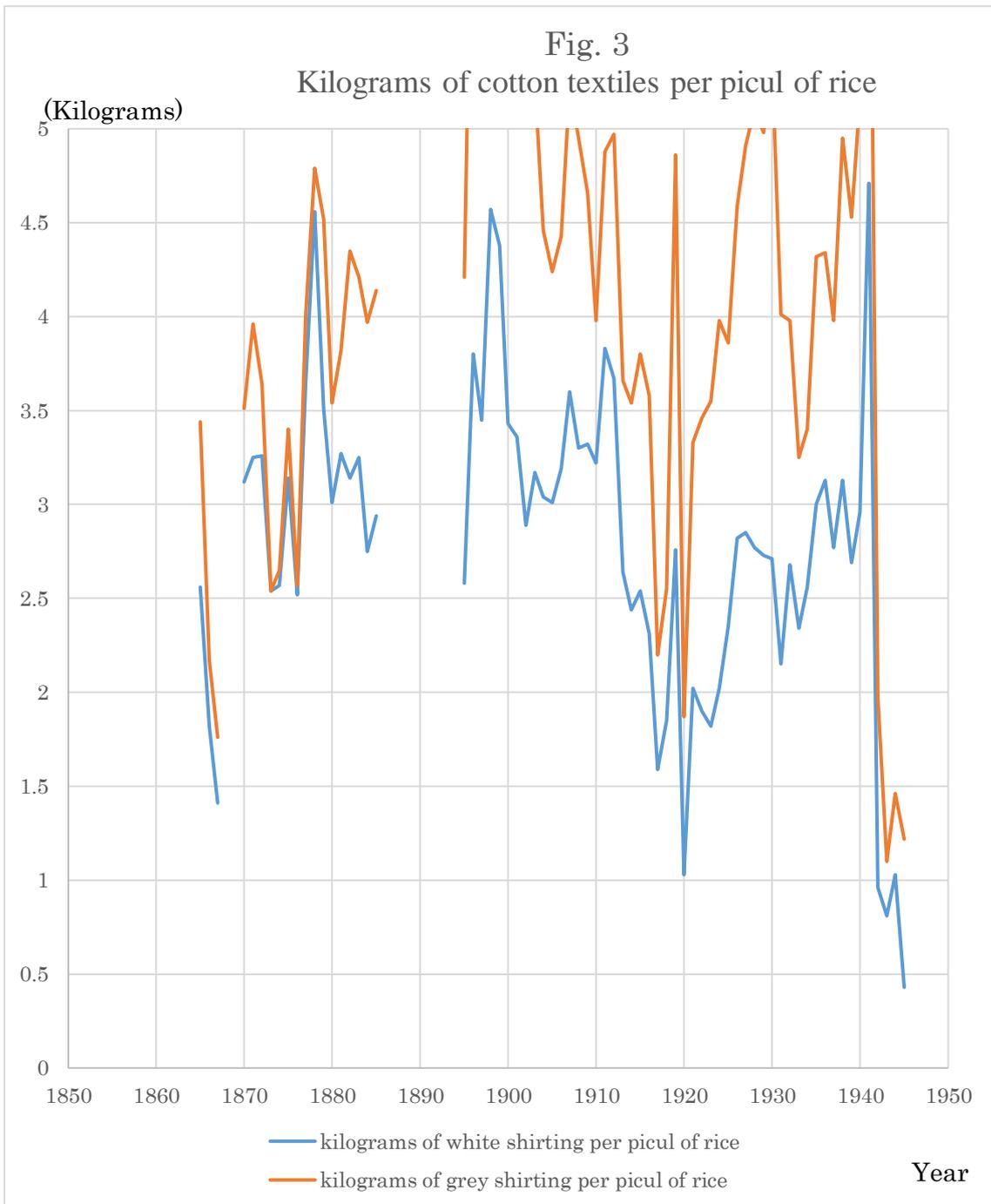

Source: Ingram (1964, p. 123, Appendix B)

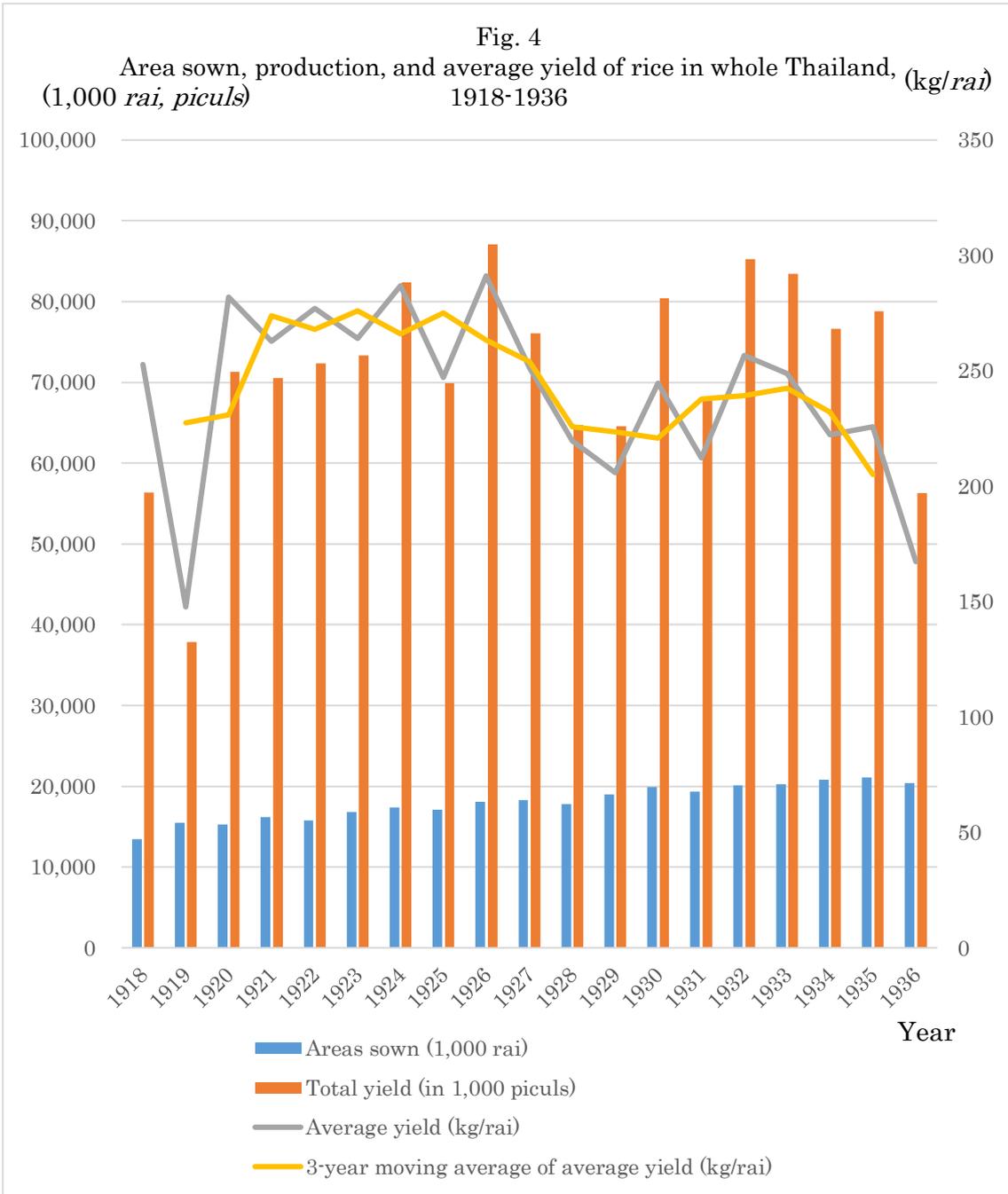

Fig. 4
Area sown, production, and average yield of rice in whole Thailand, 1918-1936

Source: SYB, No. 14 and 19.

Note: 1 *picul* = 60.48 kilogram. 6.25rai = 1hectare.

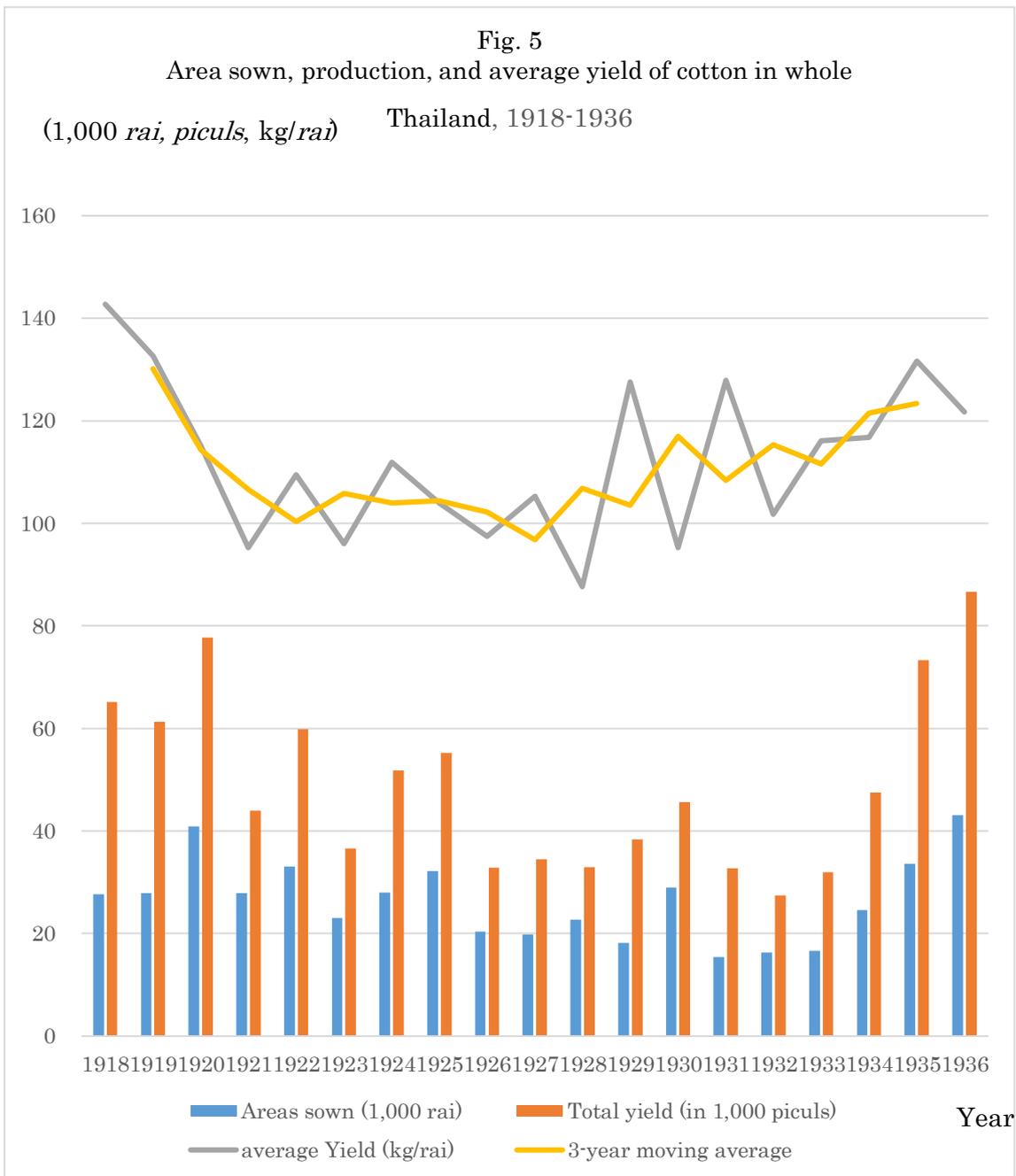

Fig. 5
Area sown, production, and average yield of cotton in whole Thailand, 1918-1936

Source: SYB, No. 14 and 19.

Note: 1 *picul* = 60.48 kilogram. 6.25rai = 1hectare.

Table 1: Area planted with 11 crops in Thailand

| No. | Item | Area sown (rai) | Exportables (E) or Importables (I) | Note |
|---|---|---|---|---|
| 1 | Rice | 18,298,440 | E | For 1927-28 |
| 2 | Tobacco | 57,515 | I | For 1927-28 |
| 3 | Maize | 50,698 | I | For 1927-28 |
| 4 | Cotton | 19,801 | I | For 1927-28 |
| 5 | Peas | 23,642 | E | For 1927-28 |
| 6 | Sesame | 11,803 | E | For 1927-28 |
| 7 | Pepper | 9,417 | E | For 1927-28 |
| 8 | Coconut | 328,945 | E | For 1927-28 |
| 9 | Fruits | 1,900,000 | I | For 1937 |
| 10 | Vegetables | 105,000 | I | For 1937 |
| 11 | Sugar | 55,835 | I | For 1924 |
| 12 | Total of 11 crops | 20,861,096 | | No. 1-11 |
| 13 | Subtotal of five exportable crops | 18,672,247 | | No.1, 5-8 |

Source: No. 1- 8 are principal crops in Thailand, SYB, No. 18 (p. 450).
No. 9 and 10 are from Ingram (1971, p. 51), and originally from Department of Agriculture, Thailand.
No. 11 is from Yamamoto (1998, p. 73).

Table B1 Comparison of the estimated total arrivals and departures of ethnic Chinese, all Thailand, annually, 1918-1934

| Year | | | (1) Skinner (1957) Arrivals | (2) Skinner (1958) Departures | (3) Skinner (1959) Net Arrivals | (4) SYB, No. 18 Arrivals | (5) SYB, No. 18 Departures | (6) SYB, No. 18 Net Arrivals | (7) Difference in Net Arrivals (3) - (6) |
|---|---|---|---|---|---|---|---|---|---|
| 1918 | - | 1919 | 67,900 | 37,000 | 30,900 | 66,901 | 36,227 | 30,674 | 226 |
| 1919 | - | 1920 | 65,700 | 43,400 | 22,300 | 64,632 | 42,684 | 21,948 | 352 |
| 1920 | - | 1921 | 70,400 | 36,800 | 33,600 | 68,797 | 35,564 | 33,233 | 367 |
| 1921 | - | 1922 | 76,500 | 46,900 | 29,600 | 73,976 | 44,967 | 29,009 | 591 |
| 1922 | - | 1923 | 95,400 | 65,200 | 30,200 | 89,329 | 60,162 | 29,167 | 1,033 |
| 1923 | - | 1924 | 115,000 | 66,400 | 48,600 | 107,987 | 60,342 | 47,645 | 955 |
| 1924 | - | 1925 | 92,700 | 66,100 | 26,600 | 84,667 | 56,258 | 28,409 | -1,809 |
| 1925 | - | 1926 | 95,500 | 60,600 | 34,900 | 86,434 | 53,112 | 33,322 | 1,578 |
| 1926 | - | 1927 | 106,400 | 73,700 | 32,700 | 100,410 | 68,744 | 31,666 | 1,034 |
| 1927 | - | 1928 | 154,600 | 76,900 | 77,700 | 139,612 | 60,791 | 78,821 | -1,121 |
| 1928 | - | 1929 | 101,100 | 72,800 | 28,300 | 88,045 | 62,138 | 25,907 | 2,393 |
| 1929 | - | 1930 | 134,100 | 68,200 | 65,900 | 70,552 | 52,677 | 17,875 | 48,025 |
| 1930 | - | 1931 | 86,400 | 62,400 | 24,000 | 76,369 | 54,219 | 22,150 | 1,850 |
| 1931 | - | 1932 | 74,800 | 56,500 | 18,300 | 69,549 | 53,058 | 16,491 | 1,809 |
| 1932 | - | 1933 | 59,500 | 44,100 | 15,400 | 51,599 | 40,627 | 10,972 | 4,428 |
| 1933 | - | 1934 | 25,700 | 32,600 | -6,900 | 15,648 | 30,176 | -14,528 | 7,628 |
| 1934 | - | 1935 | 27,000 | 31,100 | -4,100 | 24,282 | 29,305 | -5,023 | 923 |
| Total | | | 1,448,700 | 940,700 | 508,000 | 1,278,789 | 841,051 | 437,738 | 70,262 |
| Total for 1920-21 to 1926-27 | | | 651,900 | 415,700 | 236,200 | 611,600 | 379,149 | 232,451 | 3,749 |

Source: (1) - (3) are from Skinner (1957, p. 61, 173), originally based on statistics by the Immigration Division and the Customs Department in Bangkok, and given in Thailand, SYB, 1924/25-1953 and other sources. (4) - (6) are from SYB, No. 18.

Table B2 The estimated total population in each year in Thailand (1900-1950) (estimates in thousands)

| Year | (1) Est. total Population on 23 Nov. in Kobayashi (1984) | (2) Rate of change per year (%) | (3) Corrected Population on 1 Oct. in Bourgeois-Pichat (1960) | (4) Rate of change per year (%) | (5) Difference (1) - (3) |
|---|---|---|---|---|---|
| 1900 | 6845 | 1.99 | | | |
| 1901 | 6981 | | | | |
| 1902 | 7120 | | | | |
| 1903 | 7262 | | | | |
| 1904 | 7406 | | | | |
| 1905 | 7554 | | | | |
| 1906 | 7704 | | | | |
| 1907 | 7857 | | | | |
| 1908 | 8014 | | | | |
| 1909 | 8173 | | | | |
| 1910 | 8336 | | | | |
| 1911 | 8502 | | | | |
| 1912 | 8671 | | | | |
| 1913 | 8843 | | | | |
| 1914 | 9019 | | | | |
| 1915 | 9199 | | | | |
| 1916 | 9382 | | | | |
| 1917 | 9569 | | | | |
| 1918 | 9759 | | | | |
| 1919 | 9953 | | | | |
| 1920 | 10151 | | 10301 | | -150 |
| 1921 | 10353 | | 10515 | 2.08 | -162 |
| 1922 | 10559 | | 10745 | 2.19 | -186 |
| 1923 | 10770 | | 10974 | 2.13 | -204 |
| 1924 | 10984 | | 11205 | 2.10 | -221 |
| 1925 | 11170 | 2.37 | 11470 | 2.37 | -300 |
| 1926 | 11435 | | 11720 | 2.18 | -285 |
| 1927 | 11706 | | 11968 | 2.12 | -262 |
| 1928 | 11983 | | 12237 | 2.25 | -254 |
| 1929 | 12267 | | 12489 | 2.06 | -222 |
| 1930 | 12558 | | 12725 | 1.89 | -167 |
| 1931 | 12856 | | 12962 | 1.86 | -106 |
| 1932 | 13160 | | 13234 | 2.10 | -74 |
| 1933 | 13472 | | 13511 | 2.09 | -39 |
| 1934 | 13791 | | 13787 | 2.04 | 4 |
| 1935 | 14118 | | 14061 | 1.99 | 57 |
| 1936 | 14453 | | 14345 | 2.02 | 108 |
| 1937 | 14795 | | 14654 | 2.15 | 141 |
| 1938 | 15146 | | 14967 | 2.14 | 179 |
| 1939 | 15505 | | 15301 | 2.23 | 204 |
| 1940 | 15872 | | 15648 | 2.27 | 224 |
| 1941 | 16249 | | 16002 | 2.26 | 247 |
| 1942 | 16634 | | 16365 | 2.27 | 269 |
| 1943 | 17028 | | 16722 | 2.18 | 306 |
| 1944 | 17431 | | 17054 | 1.99 | 377 |
| 1945 | 17845 | | 17299 | 1.44 | 546 |
| 1946 | 18268 | | 17500 | 1.16 | 768 |
| 1947 | 18700 | | 17762 | 1.50 | 938 |
| 1948 | 19144 | | 18094 | 1.87 | 1050 |
| 1949 | 19597 | | 18477 | 2.12 | 1120 |
| 1950 | 20042 | | 18902 | 2.30 | 1140 |

Source: Kobayashi (1984, p. 56), Bourgeois-Pichat (1960, p. 25).

Table B3 Estimated total arrivals and departures (in thousands) of ethnic Chinese, all Thailand, annually and by periods, 1900–1955

| Year | | | (1) Arrivals | (2) Departures | (3) Net arrivals | (4) Total of net arrivals in each period | (5) Period | (6) Yearly average of net arrivals in each period |
|---|---|---|---|---|---|---|---|---|
| 1900 | | | 27.3 | 19.0 | 8.3 | | | |
| 1901 | | | 30.4 | 19.3 | 11.1 | | | |
| 1902 | | | 36.5 | 18.8 | 17.7 | | | |
| 1903 | | | 54.5 | 29.9 | 24.6 | | | |
| 1904 | | | 44.0 | 23.7 | 20.3 | | | |
| 1905 | | | 45.8 | 30.0 | 15.8 | | | |
| 1906 | | (1/4yr.) | 13.1 | 10.6 | 2.5 | | | |
| 1906 | – | 1907 | 68.0 | 38.9 | 29.1 | | | |
| 1907 | – | 1908 | 90.3 | 53.0 | 37.3 | | | |
| 1908 | – | 1909 | 61.6 | 49.2 | 12.4 | | | |
| 1909 | – | 1910 | 66.8 | 57.4 | 9.4 | 188.5 | 1900 to 1909–10 | 18.9 |
| 1910 | – | 1911 | 80.8 | 73.0 | 7.8 | | | |
| 1911 | – | 1912 | 76.7 | 63.9 | 12.8 | | | |
| 1912 | – | 1913 | 72.8 | 60.5 | 12.3 | | | |
| 1913 | – | 1914 | 73.3 | 57.2 | 16.1 | | | |
| 1914 | – | 1915 | 60.1 | 56.8 | 3.3 | | | |
| 1915 | – | 1916 | 69.2 | 47.1 | 22.1 | | | |
| 1916 | – | 1917 | 53.4 | 40.3 | 13.1 | | | |
| 1917 | | (3/4yr.) | 29.6 | 27.6 | 2.0 | | | |
| 1918 | | (1/4yr.) | 9.8 | 9.1 | 0.7 | | | |
| 1918 | – | 1919 | 67.9 | 37.0 | 30.9 | | | |
| 1919 | – | 1920 | 65.7 | 43.4 | 22.3 | 143.4 | 1910–11 to 1919–20 | 14.3 |
| 1920 | – | 1921 | 70.4 | 36.8 | 33.6 | | | |
| 1921 | – | 1922 | 76.5 | 46.9 | 29.6 | | | |
| 1922 | – | 1923 | 95.4 | 65.2 | 30.2 | | | |
| 1923 | – | 1924 | 115.0 | 66.4 | 48.6 | | | |
| 1924 | – | 1925 | 92.7 | 66.1 | 26.6 | | | |
| 1925 | – | 1926 | 95.5 | 60.6 | 34.9 | | | |
| 1926 | – | 1927 | 106.4 | 73.7 | 32.7 | 236.2 | 1920–21 to 1926–27 | 33.7 |
| 1927 | – | 1928 | 154.6 | 76.9 | 77.7 | | | |
| 1928 | – | 1929 | 101.1 | 72.8 | 28.3 | | | |
| 1929 | – | 1930 | 134.1 | 68.2 | 65.9 | 408.1 | 1920–21 to 1929–30 | 40.8 |
| 1930 | – | 1931 | 86.4 | 62.4 | 24.0 | | | |
| 1931 | | (3/4yr.) | 56.1 | 42.4 | 13.7 | | | |
| 1932 | | (1/4yr.) | 18.7 | 14.1 | 4.6 | | | |
| 1932 | – | 1933 | 59.5 | 44.1 | 15.4 | | | |
| 1933 | – | 1934 | 25.7 | 32.6 | −6.9 | | | |
| 1934 | – | 1935 | 27.0 | 31.1 | −4.1 | | | |
| 1935 | – | 1936 | 45.2 | 36.5 | 8.7 | | | |
| 1936 | – | 1937 | 48.9 | 28.0 | 20.9 | | | |
| 1937 | – | 1938 | 60.0 | 22.0 | 38.0 | | | |
| 1938 | – | 1939 | 33.5 | 30.0 | 3.5 | | | |
| 1939 | – | 1940 | 25.1 | 18.8 | 6.3 | 124.1 | 1930–31 to 1939–40 | 12.4 |
| 1940 | | (3/4yr.) | 23.6 | 19.8 | 3.8 | | | |
| 1941 | | | 44.8 | 36.7 | 8.1 | | | |
| 1942 | | | 11.1 | 17.8 | −6.7 | | | |
| 1943 | | | 20.1 | 20.7 | −0.6 | | | |
| 1944 | | | 18.1 | 17.9 | 0.2 | | | |
| 1945 | | | 12.4 | 11.2 | 1.2 | | | |
| 1946 | | | 86.0 | 5.8 | 80.2 | | | |
| 1947 | | | 83.8 | 23.4 | 60.4 | | | |
| 1948 | | | 28.5 | 22.3 | 6.2 | | | |
| 1949 | | | 20.0 | 15.8 | 4.2 | 157.0 | 1940–49 | 15.7 |
| 1950 | | | 7.6 | 7.4 | 0.2 | | | |
| 1951 | | | 17.9 | 13.7 | 4.2 | | | |
| 1952 | | | 9.8 | 7.3 | 2.5 | | | |
| 1953 | | | 6.4 | 2.8 | 3.6 | | | |
| 1954 | | | 4.0 | 4.5 | −0.5 | | | |
| 1955 | | | 3.8 | 4.8 | −1.0 | 9.0 | 1950–55 | 1.8 |
| Total | | | 3123.3 | 2093.2 | 1030.1 | | | |
| Total for 1900 to 1929–30 | | | 2239.3 | 1499.3 | 740.0 | | | |

Source: Skinner (1957, p. 61, 173), originally based on statistics by the Immigration Division and the Customs Department in Bangkok, and given in SYB, 1924/25–1953, and other sources.